\documentclass[a4paper,11pt]{article}
\pdfoutput=1

\usepackage{graphicx} 
\usepackage{amsmath,amsfonts,amssymb}

\usepackage{amssymb,mathrsfs,amsmath}
\usepackage{a4wide}
\usepackage{color,xcolor}
\usepackage{slashed,soul}
\usepackage[normalem]{ulem}
\usepackage{graphicx}
\usepackage{amsfonts}
\usepackage{lscape}
\def\linkcolor{cyan!70!black}

\usepackage[
colorlinks=true
,urlcolor=\linkcolor
,anchorcolor=\linkcolor
,citecolor=\linkcolor
,filecolor=\linkcolor
,linkcolor=\linkcolor
,menucolor=\linkcolor
,linktocpage=true
,pdfproducer=medialab
,pdfa=true
]{hyperref}
\usepackage{amsthm}
\usepackage{booktabs}
\usepackage{array}
\usepackage{rotating}
\usepackage[numbers, sort&compress]{natbib}
\usepackage{multirow}
\usepackage{float}
\usepackage[utf8]{inputenc}
\usepackage[T1]{fontenc}
\usepackage{appendix}
\usepackage{epsfig}
\usepackage{dsfont}

\title{Gradient Boosting MUST taggers for highly-boosted jets}

\allowdisplaybreaks

\let\OLDthebibliography\thebibliography
\renewcommand\thebibliography[1]{
  \OLDthebibliography{#1}
  \setlength{\parskip}{0pt}
  \setlength{\itemsep}{0pt plus 0.3ex}
}

\begin{document}

\begin{titlepage}

\vspace*{-3.5truecm}
\begin{flushright}
IFT-UAM/CSIC-23-29
 \end{flushright}
\vspace{0.2truecm}

\begin{center}
\renewcommand{\baselinestretch}{1.8}\normalsize
\boldmath
{\LARGE\textbf{
Gradient Boosting MUST taggers \\ for highly-boosted jets
}}
\unboldmath
\end{center}

\vspace{0.4truecm}

\renewcommand*{\thefootnote}{\fnsymbol{footnote}}

\begin{center}

{\bf 

J. A. Aguilar-Saavedra,$\,^{a}$\footnote{\href{ja.a.s@csic.es}{ja.a.s@csic.es}}
E. Arganda,$\,\,^{a,b}$\footnote{\href{ernesto.arganda@csic.es}{ernesto.arganda@csic.es}}
F. R. Joaquim,$\,\,^{c}$\footnote{\href{filipe.joaquim@tecnico.ulisboa.pt}{filipe.joaquim@tecnico.ulisboa.pt}}\\
R. M. Sand\'a Seoane$\,^{a}$\footnote{\href{mailto:r.sanda@csic.es}{r.sanda@csic.es}}
and J. F. Seabra$\,^{a,c}$\footnote{\href{joao.f.seabra@tecnico.ulisboa.pt}{joao.f.seabra@tecnico.ulisboa.pt}}}

\vspace{0.5truecm}

{\footnotesize

$^a${\sl Instituto de Física Teórica IFT UAM-CSIC, \\ C/ Nicol\'as Cabrera 13-15, Campus de Cantoblanco, 28049, Madrid, Spain \vspace{0.15truecm}}

$^b${\sl  IFLP, CONICET - Dpto. de Física, Universidad Nacional de La Plata, \\ C.C. 67, 1900 La Plata, Argentina \vspace{0.15truecm}}

$^c${\sl Departamento de Física and CFTP, Instituto Superior Técnico, Universidade de Lisboa, \\ Av.
Rovisco Pais,1, 1049-001 Lisboa, Portugal\vspace{0.15truecm}}
}

\vspace*{2mm}

\end{center}

\renewcommand*{\thefootnote}{\arabic{footnote}}
\setcounter{footnote}{0}

\begin{abstract}
\noindent 

The MUST (Mass Unspecific Supervised Tagging) method has proven to be successful in implementing generic jet taggers capable of discriminating various signals over a wide range of jet masses. We implement the MUST concept by using eXtreme Gradient Boosting ({\tt XGBoost}) classifiers instead of neural networks (NNs) as previously done. We build both fully-generic and specific multi-pronged taggers, to identify 2, 3, and/or 4-pronged signals from SM QCD background. We show that {\tt XGBoost}-based taggers are not only easier to optimize and much faster than those based in NNs, but also show quite similar performance, even when testing with signals not used in training. Therefore, they provide a quite efficient alternative machine-learning implementation for generic jet taggers.
\end{abstract}

\end{titlepage}

\section{Introduction}
\label{sec1}

Numerous searches for new physics at the Large Hadron Collider (LHC) rely on jet-tagging tools that can discriminate massive particles decaying hadronically from Quantum Chromodynamics (QCD) quark and gluon background jets. Given the importance and complexity of this task, significant efforts have been made to improve the performance of jet taggers. Following the proposal of jet substructure variables like N-subjettiness~\cite{Kim:2010uj, Thaler:2010tr, Thaler:2011gf} and energy correlation functions~\cite{Moult:2016cvt, Komiske:2017aww, Larkoski:2014gra}, several taggers using those variables to discriminate Standard Model (SM) particles, namely top quarks, Higgs bosons, or $W$/$Z$ vector bosons, have been put forward~\cite{Kim:2010uj, Thaler:2010tr, Thaler:2011gf, Moult:2016cvt, Komiske:2017aww, Larkoski:2014gra, Larkoski:2014zma, Salam:2016yht, Datta:2017rhs, CMS:2020poo}. The advent of deep learning in the last decade also opened the door to approaches employing more complex data structures like jet images~\cite{CMS:2020poo, Cogan:2014oua, Almeida:2015jua, deOliveira:2015xxd, Kasieczka:2017nvn, Lin:2018cin, Chen:2019uar} or graphs~\cite{Dreyer:2020brq, Guo:2020vvt, Gong:2022lye}.

Instead of designing tools to identify a specific signal, one can develop more generic ones, which may distinguish a broader set of multi-pronged signals from background. When trained with model-independent (MI) data, simple supervised learning algorithms can excel at that task, as shown in~\cite{Aguilar-Saavedra:2017rzt} with a shallow neural network (NN) and later in \cite{Aguilar-Saavedra:2020sxp} using logistic regression. It was also shown that these taggers can be further enhanced by applying the concept of Mass Unspecific Supervised Tagging (MUST) to their training~\cite{Aguilar-Saavedra:2020uhm}.
MUST is a method for training jet-tagging tools where the jet mass and its transverse momentum, allowed to vary over wide ranges, become input variables of a machine-learning (ML) algorithm. The results shown in~\cite{Aguilar-Saavedra:2020uhm} reveal that NNs similar to those employed in~\cite{Aguilar-Saavedra:2017rzt} can discriminate quite well any kind of multi-pronged signal over the entire ranges of jet mass and transverse momentum used in training. 
MUST jet taggers are competitive when compared to other unsupervised tools, and when integrated into a full anomaly detection method~\cite{Aguilar-Saavedra:2021utu} the performance compares well to other proposals. Quite remarkably, jet taggers developed using MUST exhibit a great resilience against modeling uncertainties~\cite{Aguilar-Saavedra:2022ejy}. And they can also be extended to identify other types of objects~\cite{Aguilar-Saavedra:2021rjk}. Despite the importance of developing jet taggers with the properties described above, there are still not many examples of tools to perform this task. A different technique to build generic taggers using supervised learning is presented in~\cite{Cheng:2022gma}. On the other hand, a complementary approach is to employ unsupervised learning. Tools to discriminate potential new physics signals from background based on unsupervised learning algorithms have also been proposed~\cite{Metodiev:2017vrx, Collins:2018epr, Collins:2019jip, Heimel:2018mkt, Farina:2018fyg, Blance:2019ibf, Hajer:2018kqm, Amram:2020ykb, Bortolato:2021zic, Cheng:2020dal, Nachman:2020lpy, Hallin:2021wme, DeSimone:2018efk, DAgnolo:2018cun, Dillon:2019cqt, Dillon:2020quc, Andreassen:2020nkr, Khosa:2020qrz}, which include some sort of jet tagging. For a review of jet substructure and different approaches to jet tagging see Ref.~\cite{Larkoski:2017jix}.

Our main goal in this work is to develop MUST taggers using eXtreme Gradient Boosting ({\tt XGBoost})~\cite{Chen:2016:XST:2939672.2939785}  which provides a scalable and highly-accurate implementation of the gradient boosting technique by pushing the limits of computing performance for boosted tree algorithms. {\tt XGBoost} has been in the spotlight for the last few years as a result of being responsible for the wins of every Kaggle data competition where companies and researchers post data for competitions on developing the best data-prediction and description models (including the winning solution of the Higgs Boson Machine Learning Challenge~\cite{Adam-Bourdarios:2015pye}). The fact that the XG Boosting algorithm uses advanced regularisation techniques to prevent overfitting and enhance its performance has also contributed to making {\tt XGBoost} one of the most popular ML tools in recent times.
In high-energy physics, {\tt XGBoost} has been applied in several different contexts including, for example, SM Higgs searches~\cite{ATLAS:2017ztq, CMS:2020tkr, ATLAS:2021ifb} and measurements~\cite{CMS:2020cga, CMS:2021sdq}, as well as searches for dark matter~\cite{ATLAS:2021jbf}, leptoquarks~\cite{ATLAS:2020xov} and lepton-flavour violation~\cite{CMS:2020kwy}.

Compared to MUST taggers previously built using NNs, the taggers we are going to present have quite similar performance in signal-to-background discrimination, while being easier to optimize and much faster  when evaluated on test sets. 
A speed improvement is an advantage of its own, especially when one also considers energy consumption. But 
this speed improvement may also be a decisive factor, for example, in the implementation of taggers in the triggering systems used by LHC experiments. The possibility of implementing top, $W$ and $Z$ jet taggers into level-1 (L1) triggers has been explored in detail in~\cite{Odagiu:2024bkp} using high-level inputs. This quite novel possibility arises from the use of particle-flow reconstruction at the L1 trigger, as planned for the CMS experiment~\cite{Zabi:2020gjd}. Jet tagging at this level would require specific hardware such as field-programmable gate arrays (FPGA) because of the extremely small latency time $O(10^{-7})$ s required in order to cope with the enormous data flow. While Ref.~\cite{Odagiu:2024bkp} only considered NNs, the implementation of gradient-boosted trees in FPGAs has already been achieved~\cite{Summers:2020xiy}, showing less than half the latency of NNs in the evaluation of track quality~\cite{CMSslides}. Another possibility is the implementation of taggers in high-level triggers, which rely on CPU farms (basically the same type of hardware that we use to obtain our results in this paper), with a required latency time $O(10^{-3})$ s. In this case, a dedicated trigger could also benefit from the improvement that {\tt XGBoost} brings in terms of computing time.

This speed improvement may also be a decisive factor to use the taggers in anomaly detection analyses over large data samples, such as the ones expected at the high luminosity phase of the LHC. In this regard, the slightly better performance of NN-based taggers would not be determinant in order to observe a potential new signal: one can conceive the use of a faster {\tt XGBoost}-based tagger in a first step of an analysis, with the use of a NN only if some significant excess is found. Last, but not least, {\tt XGBoost}-based taggers provide a different ML implementation that could be required to test the robustness of the analyses, especially if a potential new physics signal is found.

The paper is organized as follows: Section~\ref{sec2} is devoted to the presentation of the dataset used for training and the description of our {\tt XGBoost}-based tagger architecture. In Section~\ref{sec3} we show our main results. First, we study the performance of the {\tt XGBoost} taggers, comparing them with the ones previously built with NNs. Secondly, we analyze the {\tt XGBoost}-tagger capability to identify the prongness of a given signal. Finally, we show results for less generic taggers. We leave Section~\ref{sec4} to discuss our results.

\section{Training data and tagger architecture}
\label{sec2}

In this work, we use {\tt XGBoost} to build the following taggers:
\begin{itemize}
    \item a fully generic one, dubbed \texttt{XGenT}, designed to discriminate QCD jets (one-pronged, also designated by background) from multi-pronged ones (called signal);
    \item a prongness selection tagger, built for identifying the number of prongs in jets from a given sample;
    \item three multi-pronged named \texttt{XGenT$_{2P}$}, \texttt{XGenT$_{3P}$} and \texttt{XGenT$_{4P}$}, trained specifically for 2P, 3P and 4P signal discrimination.
\end{itemize}
All the data used to train the taggers is obtained from Monte Carlo simulations. Starting from the background events, we use \texttt{MadGraph}~\cite{Alwall:2014hca} in the inclusive process $pp \rightarrow jj$ to generate QCD jets. The distribution of jet mass $m_J$ is continuous for these events and we select those contained within the range $m_J\in [10,500]$~GeV. On the other hand, the generation of QCD jets is made in 100~GeV bins of transverse momentum $p_T$, covering the range $p_T\in [200,2200]$~GeV. Those bins are narrow enough to provide a smooth $p_T$ dependence, even though the events populate mostly the lower end of each bin. We should emphasize at this point that the ranges chosen for $m_J$ and $p_T$ are arbitrary and can be changed if, for instance, one intends to discriminate signals in different parameter regions from those where we test our taggers.

Regarding how multi-pronged training events are generated, we use \texttt{Protos}~\cite{protos} to generate the process $pp \rightarrow ZS$, where $S$ is a scalar and $Z \rightarrow \nu\nu$. Event samples are collected for each of the following six decay modes of $S$:
\begin{align}
    &  \text{4-pronged (4P):} && S \to u \bar u u \bar u \,,~ S \to b \bar{b} b \bar{b} \, \notag \\
    & \text{3-pronged (3P):} && S \to F \,\nu \,; \quad F \to u d d \,,~ F \to u d b \, \notag \\
    & \text{2-pronged (2P):} && S \to u \bar u \,,~ S \to b \bar b \,,
    \label{ec:MIdata}
\end{align}
where $F$ is a color-singlet fermion. In order to make the jet substructure as model-independent as possible, the decays of both $S$ and $F$ are implemented with a flat matrix element. The masses of those two particles are randomly chosen for each event within the interval $[10,800]$ GeV, with the upper limit $M_S\leq p_TR/2$ set to ensure that all decay products are contained in a jet of radius $R=0.8$. Similarly to QCD jets, all multi-pronged training samples are generated in 100~GeV bins of $p_T$. The parton-level events generated with \texttt{MadGraph} and \texttt{Protos} are processed through \texttt{Pythia}~\cite{Sjostrand:2007gs} for parton showering and hadronization, and \texttt{Delphes}~\cite{deFavereau:2013fsa} for a fast detector simulation, using the CMS card. Then, jets are reconstructed with \texttt{FastJet}~\cite{Cacciari:2011ma} applying the anti-$k_T$ algorithm~\cite{Cacciari:2008gp} with $R=0.8$, and groomed with Recursive Soft Drop~\cite{Dreyer:2018tjj}. 

For all the events described above, we obtain the mass $m_J$ and transverse momentum $p_T$ of jets, as well as 17 N-subjettiness observables,
\begin{equation}
    \left\{\tau_1^{(1/2)}, \tau_1^{(1)}, \tau_1^{(2)}, \dots , \tau_5^{(1/2)}, \tau_5^{(1)}, \tau_5^{(2)}, \tau_6^{(1)}, \tau_6^{(2)}\right\}\,,
    \label{ec:N-subjettiness}
\end{equation}
which all together characterize jet substructure. The latter are computed for ungroomed jets, following the approach proposed in~\cite{Datta:2017rhs}. That reference showed that for a four-pronged jet (which is the maximum complexity we use in training), a complete set of $\tau_N^{(\beta)}$ includes up to $\tau_3^{(2)}$, namely 8 $N$-subjettiness observables. In previous work we observed a mild improvement extending the set up to $\tau_6^{(2)}$, but higher $N$ do not bring any extra information about jet substructure.
This set of 19 variables corresponds to the inputs of our taggers. As we employ {\tt XGBoost}, a supervised learning algorithm, all jets must also be labeled according to their number of prongs and the classification task that taggers are `learning' to solve. \texttt{XGenT} and \texttt{XGenT$_{nP}$} ($n=2,3,4$) require events to be labeled either as background or signal, whereas in the prongness selection tagger, the labels identify the number of prongs.

The final step in the construction of the training datasets is to define the number of events we include in different regions of $m_J$ and $p_T$. Considering the two-dimensional space formed by those variables, we divide the $m_J$ training region, $[10,500]$ GeV, in 10 bins of 50 GeV (except the first one, within $[10,50]$ GeV) and the $p_T$ training region, $[200,2200]$ GeV, in 20 bins of 100 GeV. We discard the bins with higher jet mass but low $p_T$ due to their small number of events, keeping the full $m_J$ range just for $p_T$ bins above 1200 GeV. Inside each two-dimensional bin, we balance the number of background and signal events by including 18000 of the former and 3000 from each of the six signal jets shown in Eq.~(\ref{ec:MIdata}). Finally, we repeat the above procedure to build a validation set, which allows us to monitor the performance of {\tt XGBoost} during its training.

Some parameters of {\tt XGBoost} are tuned to optimize the Area Under the ROC Curve (AUC, with ROC standing for Receiver Operating Characteristic) obtained in the validation set. Specifically, we set for all the taggers 500 boosting trees with a maximum depth of 7 and a learning rate of 0.15. Increasing these parameters does not improve significantly the performance of the taggers. The remaining parameters of {\tt XGBoost} are set to their default values. 

In order to get our final discriminant we perform several trainings with the same datasets but different random seeds and choose the discriminant that yields the larger AUC on the validation set. The spread between AUC values among different trainings is rather small, and the ROC curves overlap. The same can be said about the NNs with which we compare. The spread between the ROC lines obtained in different trainings is minimal, and comparable to the thickness of the lines. Another method to estimate the prediction uncertainty is by using the bootstrap procedure~\cite{Efron:526679} with replacement. In every iteration of this procedure, we build a sample for training the tagger with the same size as the data set we usually use for the same end. Events are randomly picked throughout this process, allowing them to be chosen more than once. After the tagger is trained with the new training set, we determine the AUC in the validation set. (Note that this data set is the same in all the 100 iterations we perform of the bootstrap.) For the tagger \texttt{XGenT}, we have found that the size of the $95\%$ confidence level (CL) interval of the AUC is $[0.80,0.82]$.

\section{Results}
\label{sec3}

We test our taggers with several `signal' jets produced from boosted SM particles and new scalars with different masses. Some of them correspond to the multi-pronged jets for which the taggers are trained for,
\begin{itemize}
    \item 2-pronged, with: $W\rightarrow q\overline{q}$; a new scalar $A\rightarrow b\overline{b}$ with $M_A=200$~GeV;
    \item 3-pronged, with $t\rightarrow Wb\rightarrow q\overline{q}b$;
    \item 4-pronged, with: new scalars $S\rightarrow AA\rightarrow b\overline{b}b\overline{b}$, ($M_S,M_A)=(80,30)$~GeV; a new scalar $S\rightarrow WW\rightarrow q\overline{q}q\overline{q}$, $M_S=200$~GeV.
\end{itemize}
In addition, in order to test whether the taggers are capable of classifying as signal other classes of complex jets not used in training, we consider the following signals:
\begin{itemize}
    \item A heavy neutrino $N$ decaying into two quarks and a hard electron, $N\rightarrow eq\overline{q}$, with $M_N= 80$~GeV;
    \item Jets containing two quarks and two hard photons in the final state, resulting from $S\rightarrow AA\rightarrow b\overline{b}\gamma\gamma$, with ($M_S,M_A)=(200,80)$~GeV;
    \item Six-pronged jets from $A\rightarrow t\overline{t}\rightarrow q\overline{q}q\overline{q}b\overline{b}$ with $M_A = 400$ GeV.
\end{itemize}
The different types of massive jets listed above are assumed to be initiated by a heavy $Z'$ resonance with mass $M_{Z'}=1.1,$ $2.2$ or $3.3$~TeV, producing jets with transverse momenta around 0.5, 1 and 1.5 TeV, respectively.
We use in these tests a background sample generated in $pp\rightarrow jj$ with \texttt{MadGraph} in the same fashion as the one used for training and validation, and statistically independent. These samples are passed afterward through the same simulation and reconstruction chain described in Section~\ref{sec2}.

We are interested in the performance of the taggers in discriminating signal and background jets based on substructure alone. Therefore, we perform our tests in relatively narrow intervals of jet mass and $p_T$. For signals with $M_{Z'}=1.1,~2.2,~3.3$ TeV, the comparison is made in the $p_T$ intervals $[0.5, 0.6]$, $[1.0, 1.1]$ and $[1.5, 1.6]$~TeV, respectively. As for the jet mass, we consider the intervals below:
\begin{itemize}
\item $m_J\in [60,100]$ GeV for the signals with jet mass around 80 GeV: $W\rightarrow q\overline{q}$, $S\rightarrow AA\rightarrow b\overline{b}b\overline{b}$ and $N\rightarrow eq\overline{q}$;
\item $m_J\in [150,200]$ GeV for the top signal;
\item $m_J\in [160,240]$ GeV for the signals with jet mass around 200 GeV: $A\rightarrow b\overline{b}$, $S\rightarrow WW\rightarrow q\overline{q}q\overline{q}$ and $S\rightarrow AA\rightarrow b\overline{b}\gamma\gamma$.
\item $m_J\in [350,450]$ GeV for the di-top signal with jet mass around 400 GeV.
\end{itemize}

\subsection{XGBoost versus NN}
\label{sec:3.1}

\begin{figure}[p]
\begin{center}
\begin{tabular}{cc}
\includegraphics[width=0.45\textwidth]{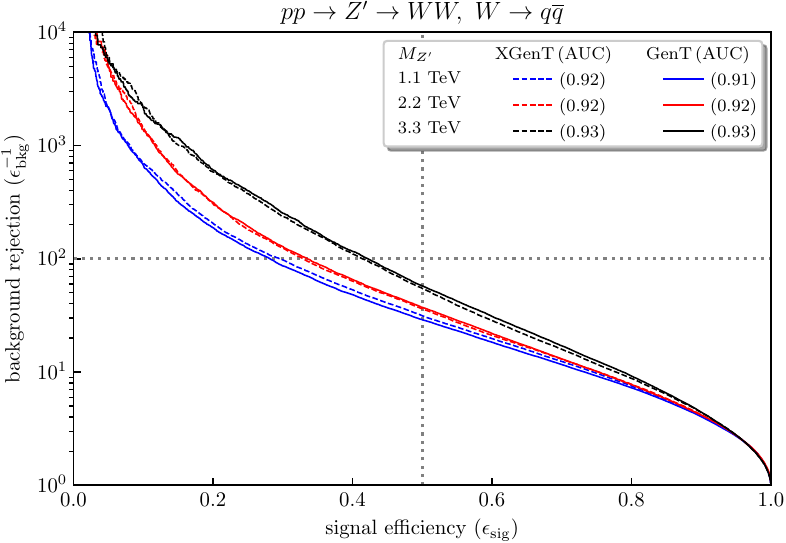} 
& \includegraphics[width=0.45\textwidth]{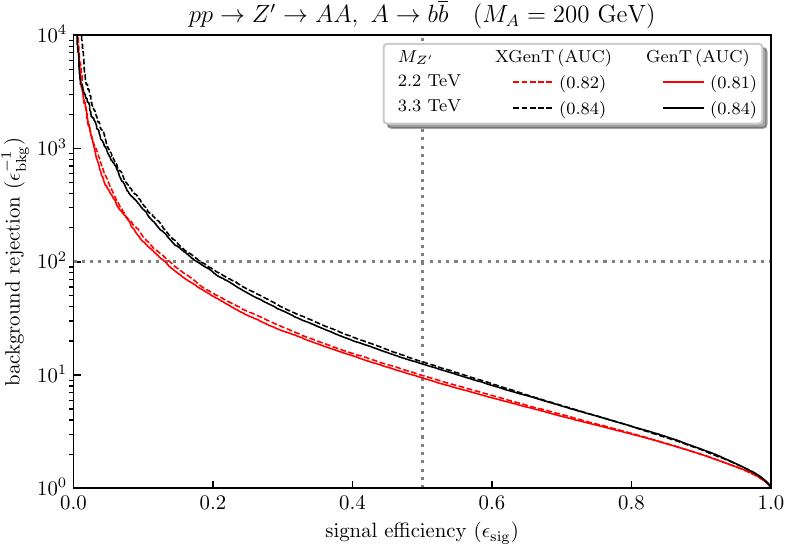}
\end{tabular}
\caption{ROC curves for \texttt{XGenT} and \texttt{GenT} on 2-pronged jet signals.}
\label{Fig:XvsNN2}
\end{center}
\end{figure}

\begin{figure}[p]
\begin{center}
\includegraphics[width=0.45\textwidth]{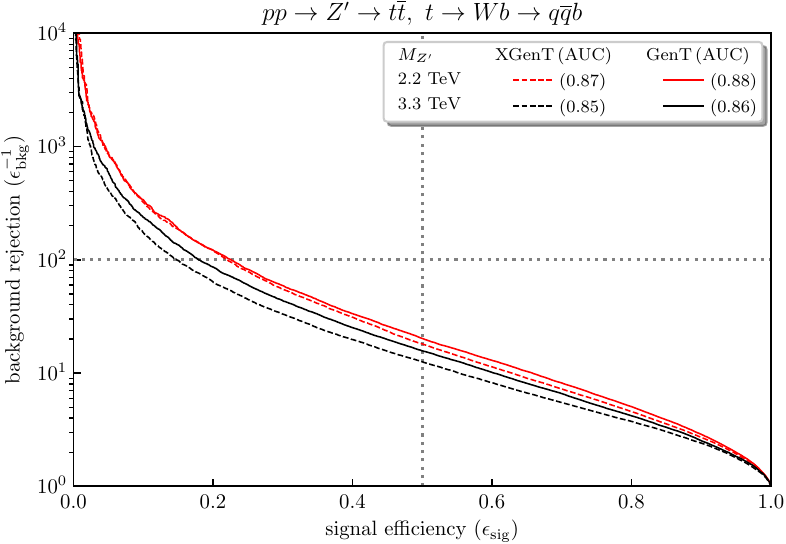} 
\caption{ROC curves for \texttt{XGenT} and \texttt{GenT} on a 3-pronged jet signal.}
\label{Fig:XvsNN3}
\end{center}
\end{figure}

\begin{figure}[p]
\begin{center}
\begin{tabular}{cc}
\includegraphics[width=0.45\textwidth]{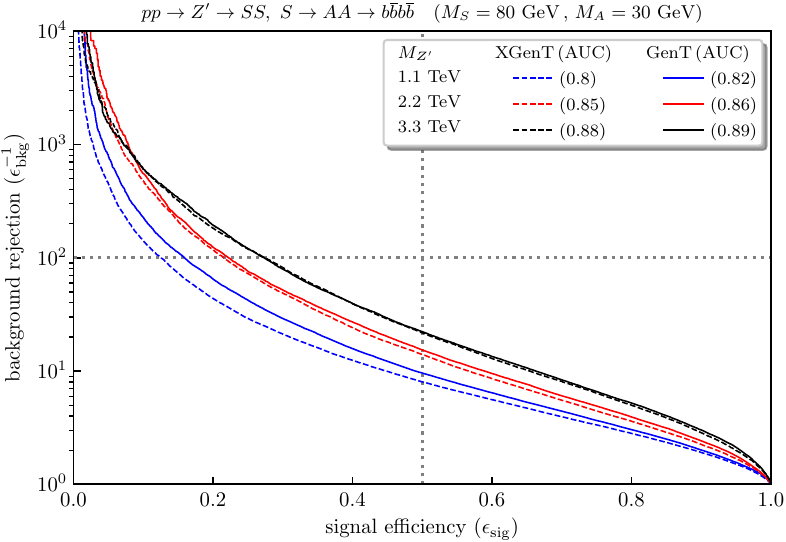} 
& \includegraphics[width=0.45\textwidth]{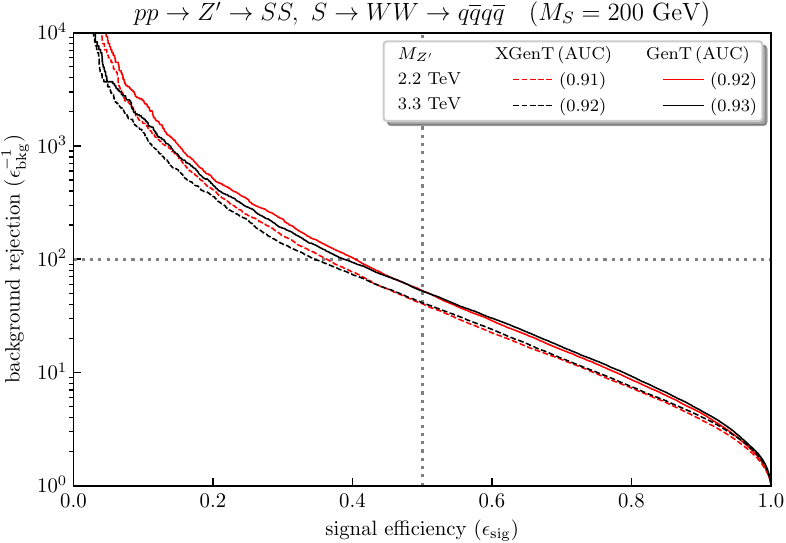} 
\end{tabular}
\caption{ROC curves for \texttt{XGenT} and \texttt{GenT} on 4-pronged jet signals.}
\label{Fig:XvsNN4}
\end{center}
\end{figure}

We first compare the performance of the fully-generic tagger \texttt{XGenT} with the performance of a fully-generic tagger \texttt{GenT} built using a NN. Figures \ref{Fig:XvsNN2}--\ref{Fig:XvsNN4} show results for two-, three- and four-pronged jets, respectively. We denote the signal and background efficiencies as $\epsilon_{\rm sig}$ and $\epsilon_{\rm bkg}$, respectively. In our plots, we include vertical lines at $\epsilon_{\rm sig} = 0.5$ and horizontal lines at $\epsilon_{\rm bkg} = 10^{-2}$ for better comparison of the performances at these benchmark values. As an additional measure of the performance, we use the AUC calculated in the $(\epsilon_{\rm sig}$, $\epsilon_{\rm bkg})$ plane.

Overall, it is seen that \texttt{XGenT} (dashed lines) and \texttt{GenT} (solid lines) have a remarkably similar performance, which is excellent in many benchmarks. The NN is found to have a better performance in four examples: top quarks with $p_T \sim 1.5$ TeV in Fig.~\ref{Fig:XvsNN3}; $S \to 4b$ with $M_S = 80$ GeV, $p_T \sim 0.5$ TeV; $S \to 4q$ with $M_S = 200$ GeV, $p_T \sim 1$ TeV and $1.5$ TeV in Fig.~\ref{Fig:XvsNN4}. These four examples in which the NN is better do not seem to follow any pattern, therefore, apparently the reason why the NN is better is simply because it can better capture the differences in substructure between the different signals and the background across very large ranges of mass and $p_T$. On the other hand, the different performance found for the different benchmarks arises, as already discussed in Ref.~\cite{Aguilar-Saavedra:2020uhm}, due to the substructure of the jets at the particular mass and $p_T$ range considered. Notice in particular that QCD jets acquire a large mass due to extra radiation, in which case their substructure is not actually one-pronged. 

For the signals in which the NN performs better, the difference is not determinant. Let us consider $S \to AA \to b \bar b b \bar b$ with $M_S = 80$ GeV, $p_T \sim 0.5$ TeV, for which the difference is larger. An experimental search will set a threshold on background rejection, say $\epsilon_\text{bkg}^{-1} = 10^2$. For this rejection, the signal efficiency is 1.3 times higher for the NN. In a search with two tags, the signal significance will be 1.7 times larger. This factor, though important, is not likely to be determinant for the observability of a possible signal. And, as previously discussed, a quick scan for anomalies with  {\tt XGBoost} could be complemented by a NN in the regions of interest.

\begin{figure}[t]
\begin{center}
\includegraphics[width=0.45\textwidth]{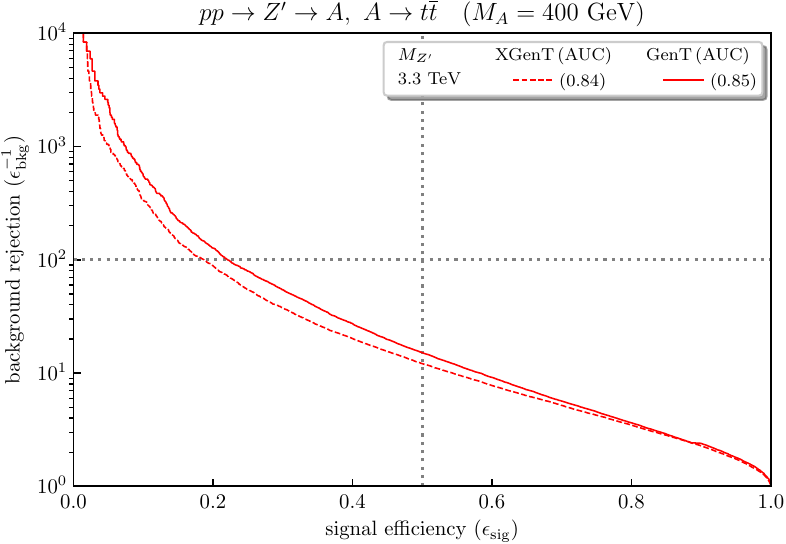}
\caption{ROC curves for \texttt{XGenT} and \texttt{GenT} on a 6-pronged jet signal.}
\label{Fig:XvsNN6}
\end{center}
\end{figure}

\begin{figure}[t]
\begin{center}
\begin{tabular}{cc}
\includegraphics[width=0.45\textwidth]{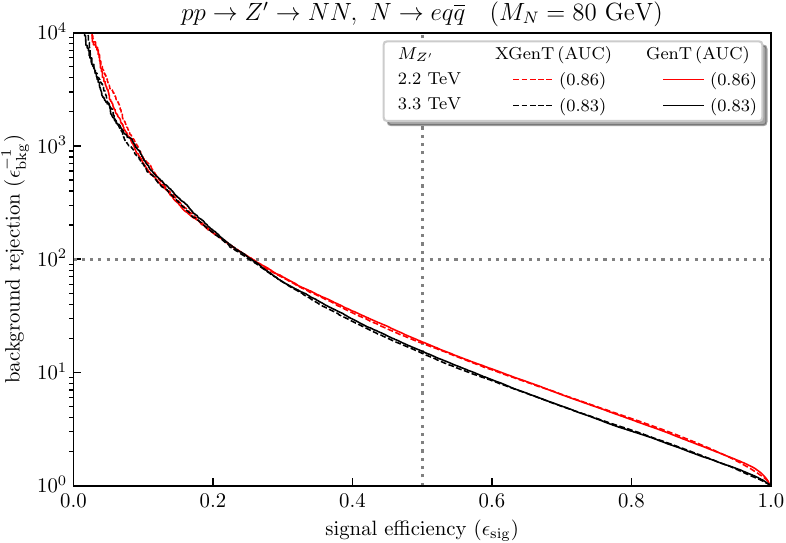} 
& \includegraphics[width=0.45\textwidth]{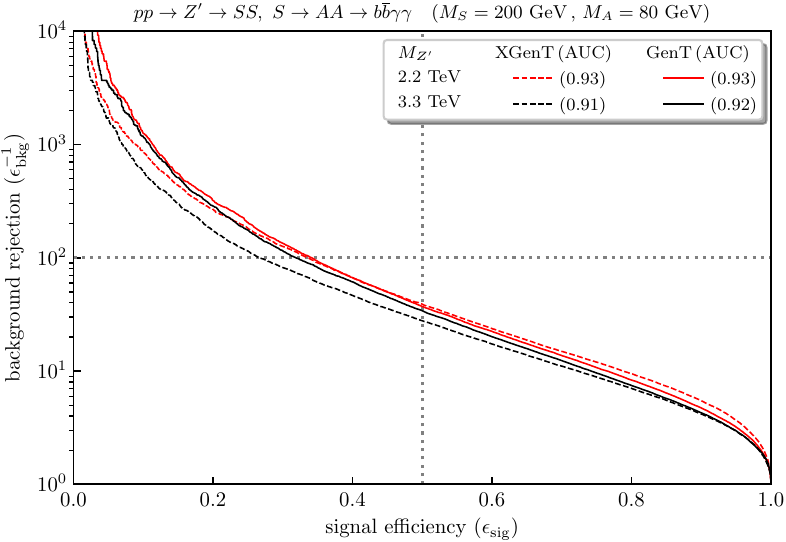} 
\end{tabular}
\caption{ROC curves for \texttt{XGenT} and \texttt{GenT} on jet signals with electrons or photons}
\label{Fig:XvsNNx}
\end{center}
\end{figure}

For signal jets with a topology different from those used in training, the conclusion is similar. Figure~\ref{Fig:XvsNN6} shows results for six-pronged di-top jets, and Fig.~\ref{Fig:XvsNNx} for jets containing either a hard electron or hard photons. In some cases \texttt{GenT} is slightly better, but without any apparent pattern.

\subsection{Signal identification}
\label{sec:3.2}

The preceding results show that the generic tagger \texttt{XGenT} can efficiently discriminate multi-pronged jets produced by boosted particles from background. However, this tagger does not provide any information regarding the prongness of the signal, which might be required to identify the particle producing it. This can be achieved with a different type of tagger that classifies jets by prongness. Such a tool was investigated in Ref.~\cite{Aguilar-Saavedra:2020uhm} using NNs and we now show that a similar tagger can be built using {\tt XGBoost}.

The training sample is the same as the one employed for \texttt{XGenT}, containing background, 2P, 3P, and 4P jets. We note that this tagger does not perform a binary classification task (like \texttt{XGenT}) but instead uses a different label for each of the four types of jets. Thus, the output of {\tt XGBoost} provides the probabilities $P_{\rm bkg}$, $P_{2P}$, $P_{3P}$ and $P_{4P}$ for an event being, respectively, background, two-, three- or four-pronged. The metric we optimize during training is accuracy, i.e. the percentage of events correctly classified by the tagger. 

We test our prongness selection tagger in five benchmarks selected from the ones considered before:
\begin{itemize}
    \item 2P1: $W$ bosons in $W \to q \bar q$, with $M_{Z'}=2200$~GeV;
    \item 2P2: New scalars $A \rightarrow b\overline{b}$ with $M_A=200$~GeV and $M_{Z'}=3300$~GeV;
    \item 3P: Top quarks in $t\rightarrow q \bar q b$ with $M_{Z'}=2200$~GeV;
    \item 4P1: New scalars $S\rightarrow AA\rightarrow b \bar b b \bar b$ with $M_S=80$~GeV, $M_A=30$~GeV, $M_{Z'}=3300$~GeV; 
    \item 4P2: New scalars $S\rightarrow WW\rightarrow q \bar q q \bar q$ with $M_S=200$~GeV, $M_{Z'}=2200$~GeV.
\end{itemize}
The first and second benchmarks correspond to two-pronged jets, in the third one the jets are three-pronged, and in the last two benchmarks they are four-pronged.
In all these benchmarks we require $P_{\rm sel} \equiv \max(P_{\rm bkg}, P_{2P}, P_{3P}, P_{4P})$ to be larger than a threshold $P_{\rm min}$ for the events to be considered as correctly classified. While this improves the accuracy of the tagger, it also implies that not all events can be assigned to a class, if that requirement is not fulfilled. The events that are not assigned to either class form a new class of events called \emph{undefined}, which contains all those with $P_{\rm sel} < P_{\rm min}$.

\begin{figure}[t]
\begin{center}
\includegraphics[width=\textwidth]{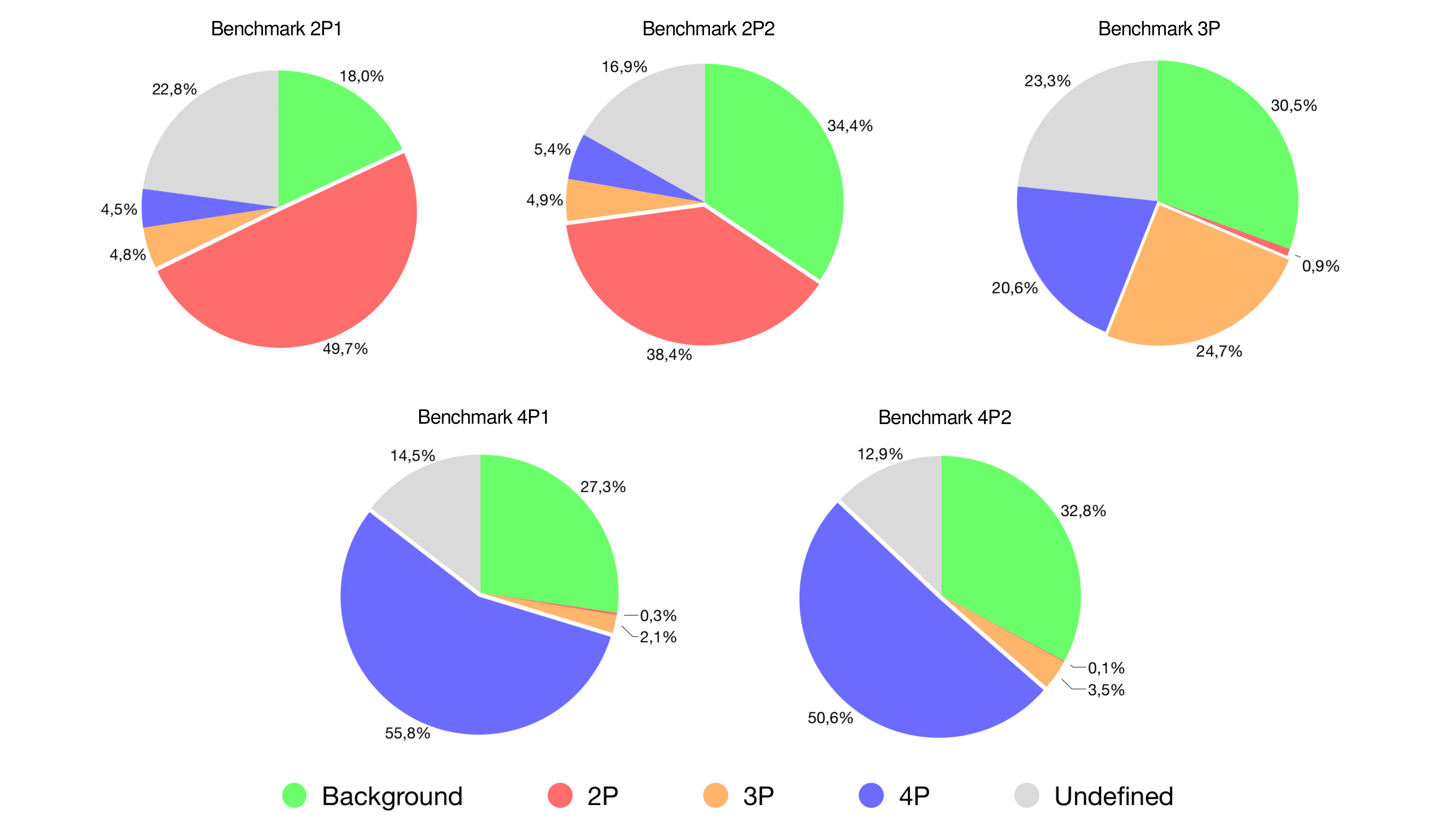}
\caption{Fraction of jets classified as Background, 2P, 3P, 4P, or Undefined for the five benchmarks described in the text.}
\label{Fig:PieCharts}
\end{center}
\end{figure}

Considering $P_{\rm min} = 0.4$, we show in Fig.~\ref{Fig:PieCharts} the fraction of events classified as Background, 2P, 3P, 4P, and undefined in each benchmark.
 In all cases except for boosted tops, the class that has a larger fraction of events corresponds to the true event class. It is also seen that four-pronged jets are correctly classified more than 50\% of the time. Notably, a large fraction of signal events are identified as background. This happens, as aforementioned, because QCD jets with a large mass (which are very rare, anyway) arise from the radiation of a quark or gluon. In that case, their substructure is multi-pronged, the same as for the signal.

\begin{figure}[t!]
\begin{center}
\begin{tabular}{cc}
\includegraphics[keepaspectratio, width=0.45\textwidth]{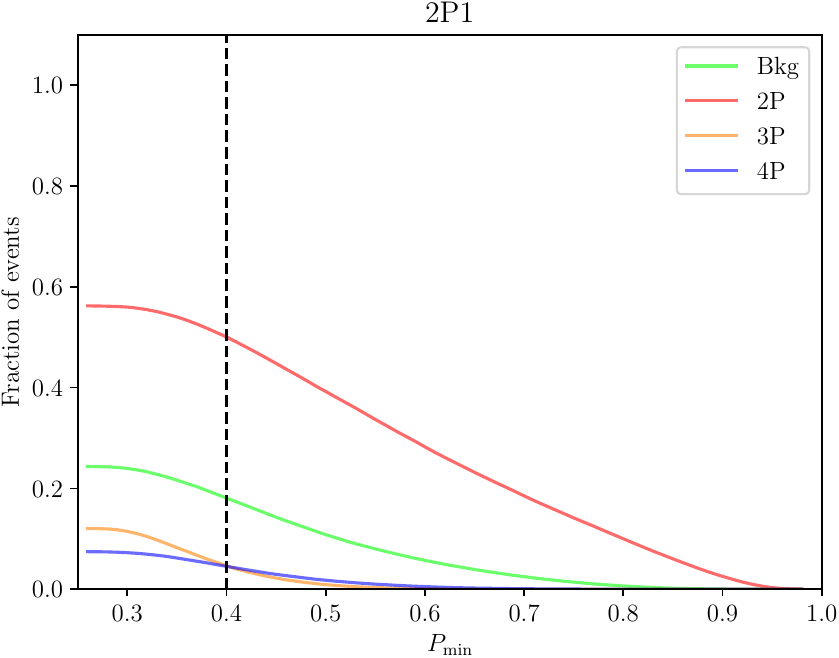} &
\includegraphics[keepaspectratio, width=0.45\textwidth]{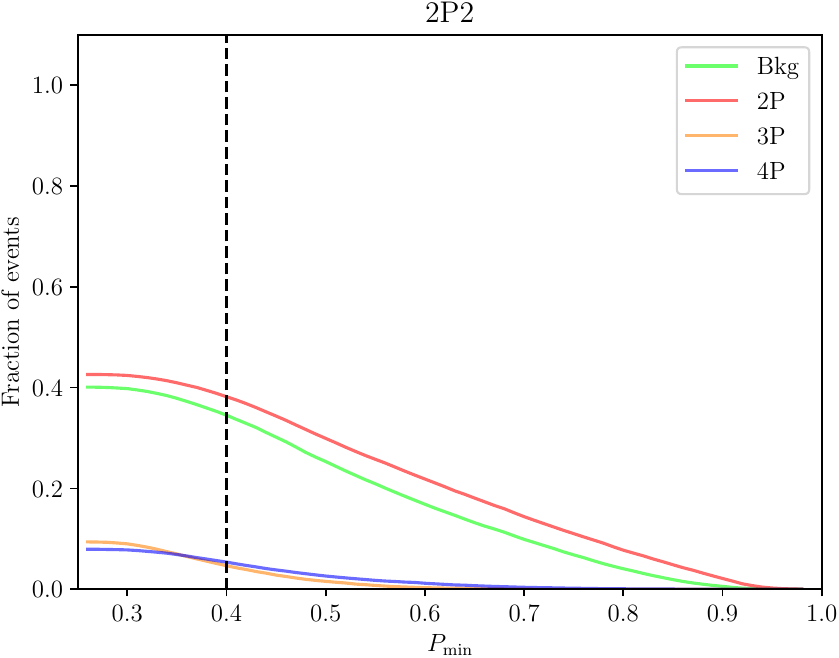} \\[1mm]
\includegraphics[keepaspectratio, width=0.45\textwidth]{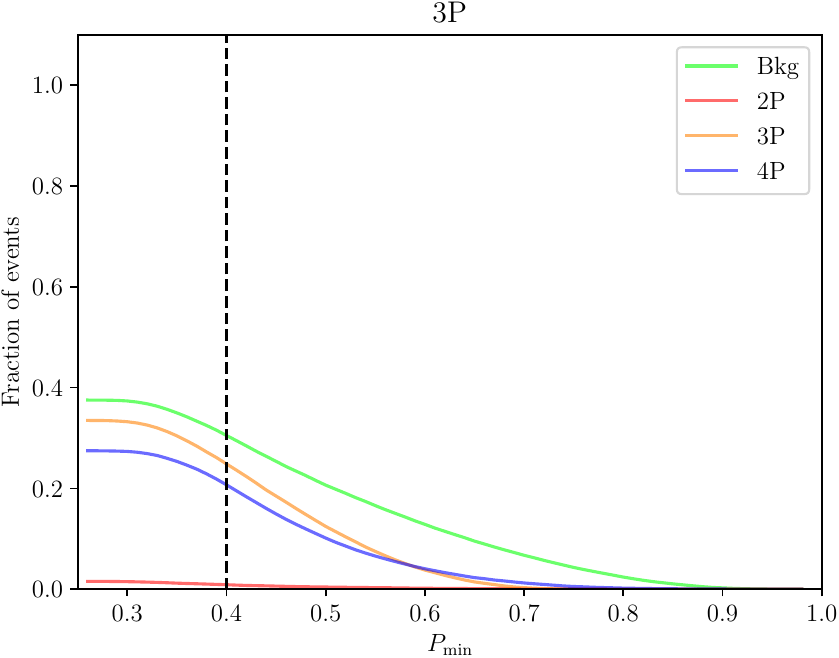} &
\includegraphics[keepaspectratio, width=0.45\textwidth]{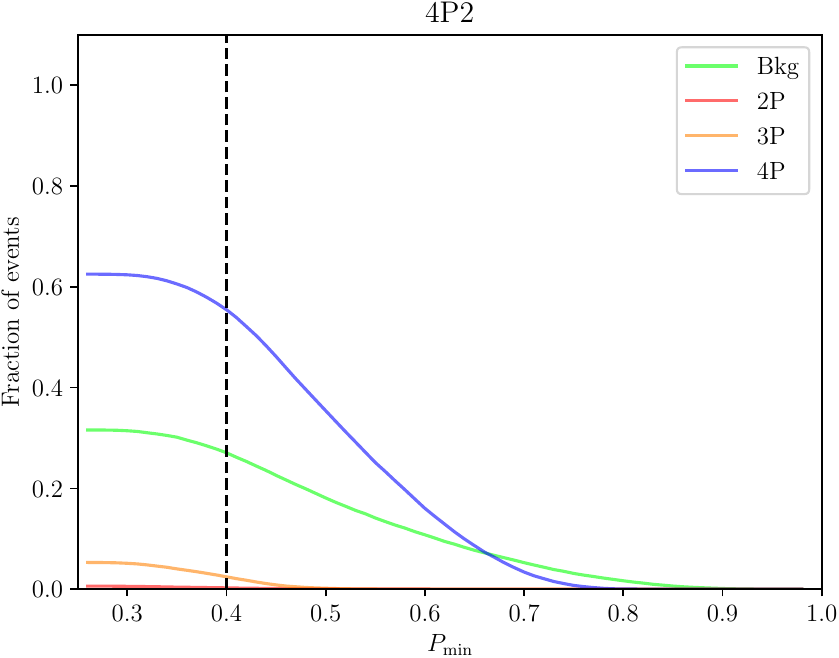} \\[1mm]
\multicolumn{2}{c}{\includegraphics[keepaspectratio, width=0.45\textwidth]{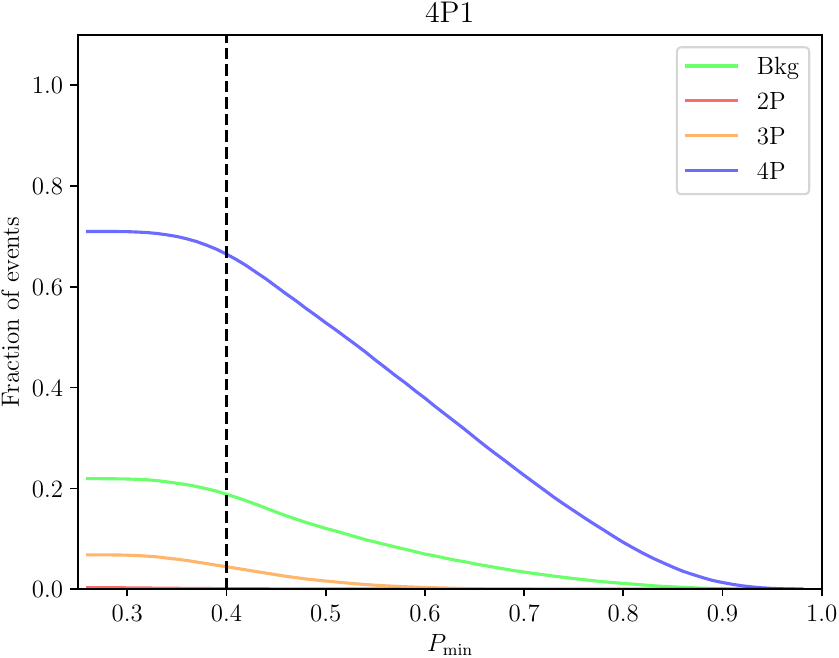}}
\end{tabular}
\end{center}
\caption{Fraction of events classified as Background, 2P, 3P and 4P as a function of $P_{\rm min}$. The dashed line represents our choice of $P_{\rm min} = 4$ to obtain the results of Fig.~\ref{Fig:PieCharts}.}
\label{Fig:ThrsVsPerc}
\end{figure}

In all benchmarks, the threshold $P_{\rm min} = 0.4$ leaves a fairly small fraction of events in the undefined class. Other choices of $P_{\rm min}$ are explored in Fig.~\ref{Fig:ThrsVsPerc}. In these plots we show for each benchmark the fraction of events classified as background, two, three and four-pronged, for each choice of $P_{\rm min}$. Note that in the presence of four classes the minimum threshold we may define is $P_{\rm min} = 0.25$, and as $P_{\rm min} \to 1$ the number of events with $P_{\rm sel} \geq  P_{\rm min}$ in any class goes to zero, thereby enlarging the undefined class. From these plots we can also observe that for any $P_{\rm min}$ in the $0.25-0.5$ ballpark the results are equivalent.

\begin{table}[t]
\centering 
\setlength{\tabcolsep}{5pt}
\renewcommand{\arraystretch}{1.25}
\begin{tabular}{cccccc} \toprule 
\multirow[c]{2}{*}{benchmark} & \multicolumn{5}{c}{Classifier output} \\[0.5em]
& background & 2P & 3P & 4P & undefined \\
\midrule
2P1 & [18.4, 21.1] & [47.1, 50.8] & [5.4, 7.8] & [4.7, 7.4] & [16.2, 22.7]  \\
2P2 & [33.7, 37.2] & [35.9, 39.9] & [5.0, 8.0] & [5.1, 6.8] & [12.8, 17.2] \\
3P & [29.6, 32.9] & [0.9, 1.3] & [25.1, 33.6] & [18.1, 20.7] & [14.4, 23.2]  \\ 
4P1 & [27.4, 32.1] & [0.2, 0.7] & [2.1, 7.3] & [48.8, 55.3] & [10.2, 15.9]  \\
4P2 & [18.3, 22.5] & [0.1, 0.2] & [4.8, 8.1] & [61.5, 66.3] & [7.6, 10.6] \\
\bottomrule
\end{tabular}
\caption{95\% CL interval for the percentage of events classified as background, 2P, 3P, 4P and undefined, estimated with the bootstrap method for $P_{\rm min} = 0.4$. The central values correspond to those in fig.~\ref{Fig:ThrsVsPerc}.}
\label{tab:unc}
\end{table}

The uncertainty in the predictions can be estimated with the bootstrap method, and collected in Table~\ref{tab:unc}, for the full signal datasets. For small ``signal'' datasets, the uncertainties are larger, reflecting the fact that the  identification of the prongness of a possible signal, discriminating among various hypotheses, requires a sizeable sample. To illustrate this fact, we select four events from benchmark 4P1, those that have the highest value of $P_\text{bkg}$, $P_\text{2P}$, $P_\text{3P}$ and $P_\text{4P}$. The 95\% CL uncertainties on these probabilities are
\begin{align}
& \text{event \#1:} \quad P_\text{bkg} \in [89.4,97.4] \,, \notag \\
& \text{event \#2:} \quad P_\text{2P} \in [45.2,75.4] \,, \notag \\
& \text{event \#3:} \quad P_\text{3P} \in [40.9,73.4] \,, \notag \\
& \text{event \#4:} \quad P_\text{4P} \in [61.2,84.4] \,.
\end{align}

\subsection{Taggers for specific prongness}

\begin{figure}[b]
\begin{center}
\begin{tabular}{cc}
\includegraphics[width=0.45\textwidth]{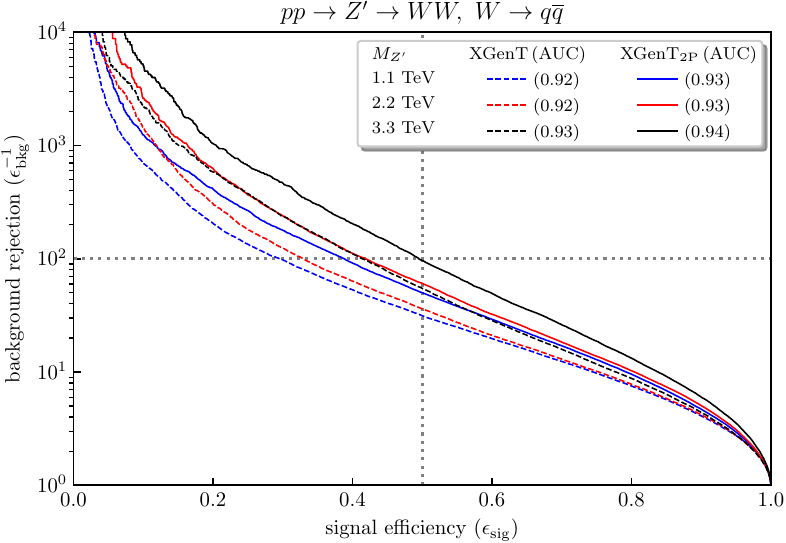} 
& \includegraphics[width=0.45\textwidth]{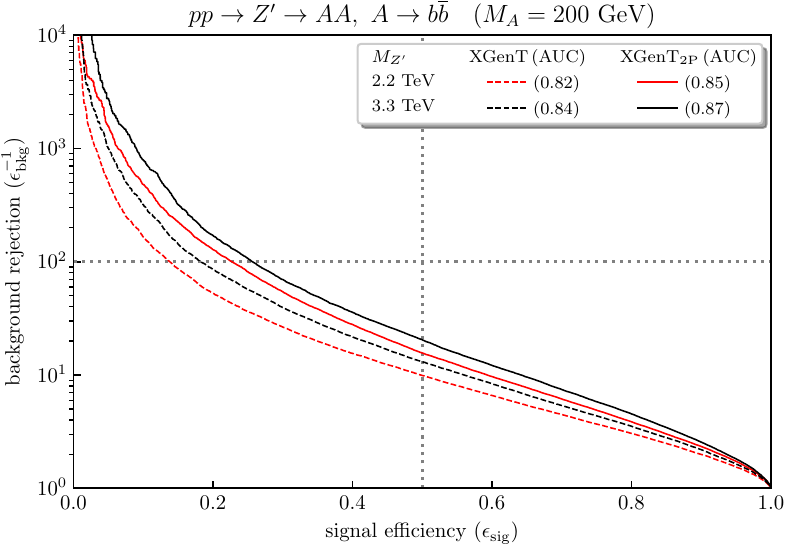} \\
\end{tabular}
\caption{\texttt{XGenT} vs. \texttt{XGenT$_{2P}$} for 2-pronged jet signals.}
\label{Fig:2P}
\end{center}
\end{figure}

\begin{figure}[tb]
\begin{center}
\includegraphics[width=0.45\textwidth]{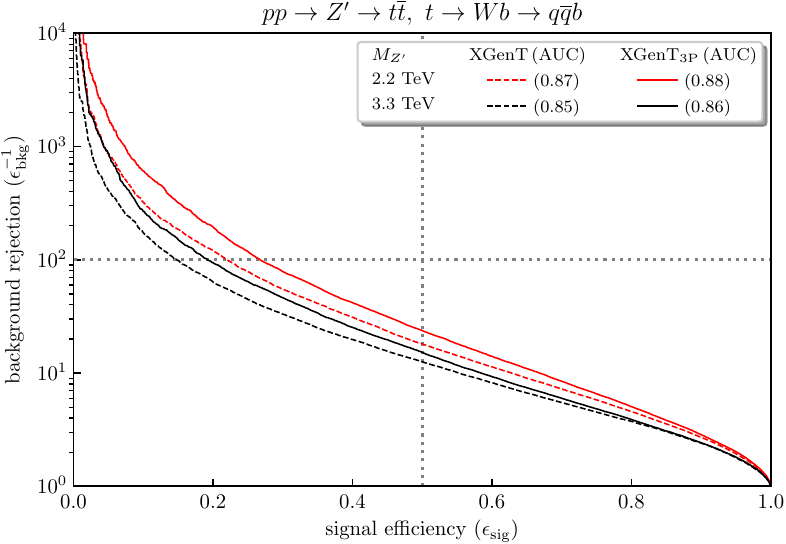}
\caption{\texttt{XGenT} vs. \texttt{XGenT$_{3P}$} for a 3-pronged jet signal.}
\label{Fig:3P}
\end{center}
\end{figure}

\begin{figure}[tb]
\begin{center}
\begin{tabular}{cc}
\includegraphics[width=0.45\textwidth]{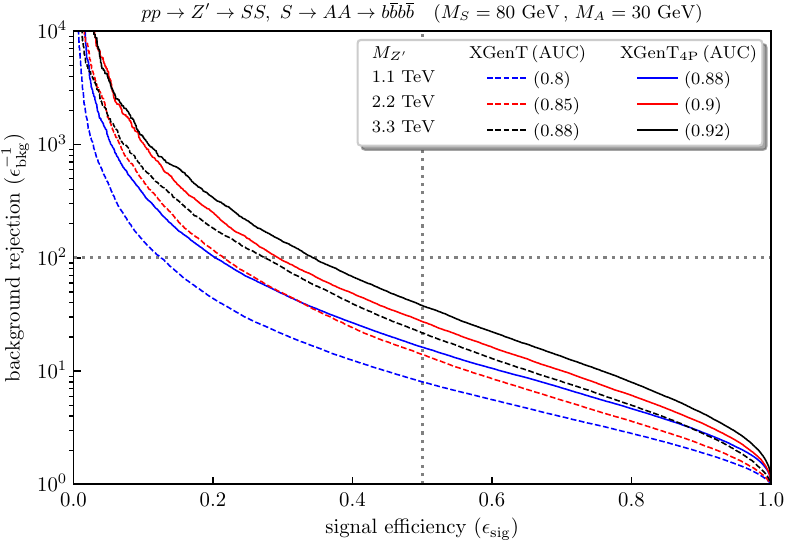} 
& \includegraphics[width=0.45\textwidth]{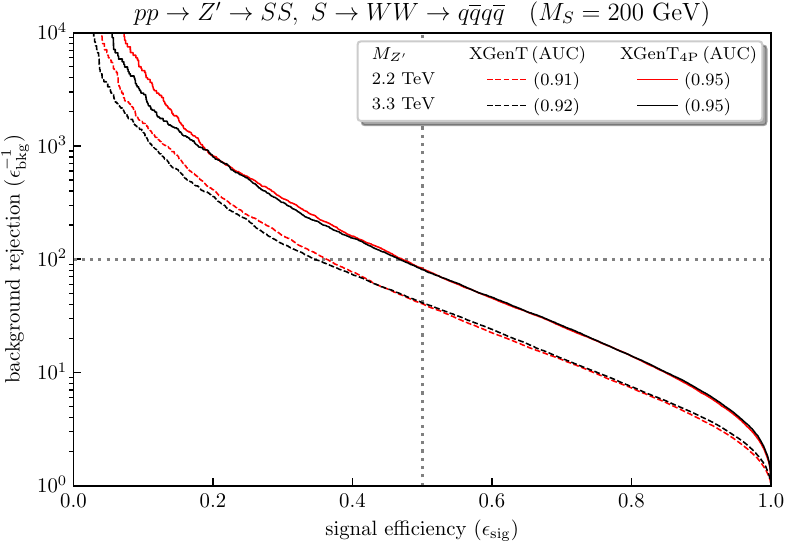}
\end{tabular}
\caption{\texttt{XGenT} vs. \texttt{XGenT$_{4P}$} for 4-pronged jet signals.}
\label{Fig:4P}
\end{center}
\end{figure}

In some applications, it may be useful to have taggers trained to discriminate massive jets of a given prongness (2P, 3P, or 4P) from the background. Though less general, these taggers generally have a better performance when applied to the signal jets designed for, as it is expected (and in agreement with results in Ref.~\cite{Aguilar-Saavedra:2020uhm}). For completeness, we compare these taggers, which we label as \texttt{XGenT$_{2P}$}, \texttt{XGenT$_{3P}$} and \texttt{XGenT$_{4P}$}, respectively with the fully-generic tagger \texttt{XGenT}. Figures~\ref{Fig:2P}--\ref{Fig:4P} show the results for two-, three- and four-pronged jets.


\section{Discussion}
\label{sec4}

Along this work, we have applied the MUST concept to develop jet taggers using {\tt XGBoost} classifiers. We have built a fully generic \texttt{XGenT} tagger to discriminate one-pronged QCD jets (our background) from multi-pronged ones (our signal); a prongness selection tagger to identify the number of prongs in jets from a given sample; and three multi-pronged taggers to discriminate 2P, 3P and 4P signals from background. These taggers have been tested with various signals, leading to an excellent signal to background separation in most cases. 

The performance of the fully-generic tagger \texttt{XGenT} is quite close to that of its NN counterpart \texttt{GenT} used in Refs.~\cite{Aguilar-Saavedra:2020uhm,Aguilar-Saavedra:2021utu}, not only on multi-pronged jets as used in the training but also on other types of complex jets. This is an interesting result by itself since the number of generic jet taggers available in the literature is somewhat scarce. The use of alternative methods to look for new physics on data is required to test the robustness of the searches and is a must in case any excess over the SM expectation is found. 

In addition, we have found that \texttt{XGenT} is much faster than the NN version. In applications where jet tagging is intensive and the speed is critical (e.g. in anomaly detection tools) \texttt{XGenT} provides a significant advantage. Table~\ref{tab:CPUtime} collects the time spent to train \texttt{XGenT} and \texttt{GenT} (first row), as well as the time to apply the two taggers to the five signals that we studied in both sections~\ref{sec:3.1} and \ref{sec:3.2} (second to sixth rows). Those signals are labelled in the same way as the benchmarks in section~\ref{sec:3.2}. The computation times, measured using an Intel Core i9-9900K CPU, are proportional to the number of events being tagged. No relation between the type of signal being tagged and the computation time was found in our measurement. These times are still well above the latency times of L1 or high-level triggers, indicating that faster CPUs or parallel processing would be required to implement our MUST generic tagger into a trigger system.

\begin{table}[h]
\centering
\setlength{\tabcolsep}{15pt}
\renewcommand{\arraystretch}{1.25}
\begin{tabular}{cccc} \toprule 
Dataset & Number of events & \texttt{XGenT} (s) & \texttt{GenT} (s) \\
\midrule
Training & 5594624 & 730.33 & 2520.81 \\
Signal - 2P1 & 274513 & 0.21 & 1.81 \\
Signal - 2P2 & 136579 & 0.11 & 0.91 \\
Signal - 3P & 151875 & 0.12 & 1.02 \\ 
Signal - 4P1 & 166480 & 0.13 & 1.10 \\
Signal - 4P2 & 181579 & 0.14 & 1.21 \\
\bottomrule
\end{tabular}
\caption{Computing time (third and fourth column) for training and tagging events using XGBoost and a NN. The signals shown in this table are labelled as in Section~\ref{sec:3.2}.}
\label{tab:CPUtime}
\end{table}

The related task of identifying a possible signal has also been addressed here with a {\tt XGBoost} tagger that is trained to classify jets by prongness. The accuracy of the classification is good for all new physics signals from new scalars (two- and four-pronged), as well as for SM $W$ jets. The classification only fails for boosted top quarks, which are often classified as two- or four-pronged. This, however, is not a relevant issue, as possible new physics signals with boosted tops will have other signatures (with top semileptonic decays) that allow identifying the top quarks more cleanly than a jet tagger could possibly do. (This statement also applies to signals with boosted $W$ bosons, which can be better identified by their leptonic decay.)

Finally, for completeness, we have investigated other types of less generic taggers, in which only jets of a given prongness are taken as signal. Though these taggers are quite more specific, they also have better performance for the signal they are trained for and might be useful in some cases.

In conclusion, the MUST taggers developed using XGBoost are a viable alternative to those based in NNs, and constitute a useful addition to the jet tagging toolset, that may be used by LHC experiments in generic searches for new physics.


\section*{Acknowledgments}
EA and RMSS would like to thank Diego Mouriño for his collaboration with code. This work is partially supported by the ``Atracci\'on de Talento'' program (Modalidad 1) of the Comunidad de Madrid (Spain) under the grant number 2019-T1/TIC-14019 (EA, RMSS), by the Spanish Research Agency (Agencia Estatal de Investigaci\'on) through the Grant IFT Centro de Excelencia Severo Ochoa No CEX2020-001007-S (JAAS, EA, RMSS) and by the grants PID2019-110058GB-C21, PID2022-142545NB-C21 (JAAS) and PID2021-124704NB-I00 (EA, RMSS) funded by MCIN/AEI/10.13039/501100011033. FRJ and JFS acknowledge financial support from Fundação para a Ciência e a Tecnologia (FCT, Portugal) through the projects CFTP-FCT Unit (UIDB/00777/2020, UIDP/00777/2020) and CERN/FIS-PAR/0019/2021, which are partially funded through POCTI (FEDER), COMPETE, QREN, and EU. The work of J.F.S. is supported by the FCT grant SFRH/BD/143891/ 2019.

\section*{Data availability statement}

Parton-level MC samples used in this work are available at https://jaguilar.web.cern.ch/jaguilar/ multiprong/


\begin{thebibliography}{99}

\bibitem{Kim:2010uj}
J.-H.~Kim,
Phys. Rev. D \textbf{83}, 011502 (2011)
[arXiv:1011.1493 [hep-ph]].

\bibitem{Thaler:2010tr}
J.~Thaler and K.~V.~Tilburg,
JHEP \textbf{03}, 015 (2011)
[arXiv:1011.2268 [hep-ph]].

\bibitem{Thaler:2011gf}
J.~Thaler and K.~V.~Tilburg,
JHEP \textbf{02}, 093 (2012)
[arXiv:1108.2701 [hep-ph]].

\bibitem{Moult:2016cvt}
I.~Moult and L.~Necib and J.~Thaler,
JHEP \textbf{12}, 153 (2016)
[arXiv:1609.07483 [hep-ph]].

\bibitem{Komiske:2017aww}
P.~T.~Komiske and E.~M.~Metodiev and J.~Thaler,
JHEP \textbf{04}, 013 (2018)
[arXiv:1712.07124 [hep-ph]].

\bibitem{Larkoski:2014gra}
A.~J.~Larkoski and I.~Moult and D.~Neill,
JHEP \textbf{12}, 009 (2014)
[arXiv:1409.6298 [hep-ph]].

\bibitem{Larkoski:2014zma}
A.~J.~Larkoski and I.~Moult and D.~Neill,
Phys. Rev. D \textbf{91}, no. 3, 034035 (2015)
[arXiv:1411.0665 [hep-ph]].

\bibitem{Salam:2016yht}
G.~P.~Salam and L.~Schunk and G.~Soyez,
JHEP \textbf{03}, 022 (2017)
[arXiv:1612.03917 [hep-ph]].

\bibitem{Datta:2017rhs}
K.~Datta and A.~Larkoski,
JHEP \textbf{06}, 073 (2017)
[arXiv:1704.08249 [hep-ph]].

\bibitem{CMS:2020poo}
A.~Sirunyan \emph{et al.} [CMS],
JINST \textbf{15}, no. 6, P06005 (2020)
[arXiv:2004.08262 [hep-ex]].

\bibitem{Cogan:2014oua}
J.~Cogan, M.~Kagan, E.~Strauss and A.~Schwartzman,
JHEP \textbf{02}, 118 (2015)
[arXiv:1407.5675 [hep-ph]].

\bibitem{Almeida:2015jua}
L.~G.~Almeida, M.~Backovi\'c, M.~Cliche and S.~J.~Lee and M.~Perelstein,
JHEP \textbf{07}, 086 (2015)
[arXiv:1501.05968 [hep-ph]].

\bibitem{deOliveira:2015xxd}
L.~de~Oliveira, M.~Kagan, L.~Mackey, B.~Nachman and A.~Schwartzman,
JHEP \textbf{07}, 069 (2016)
[arXiv:1511.05190 [hep-ph]].

\bibitem{Kasieczka:2017nvn}
G.~Kasieczka, T.~Plehn, M.~Russell and T.~Schell,
JHEP \textbf{05}, 006 (2017)
[arXiv:1701.08784 [hep-ph]].

\bibitem{Lin:2018cin}
J.~Lin, M.~Freytsis, I.~Moult and B.~Nachman,
JHEP \textbf{10}, 101 (2018)
[arXiv:1807.10768 [hep-ph]].

\bibitem{Chen:2019uar}
J.~Y.-C.~Chen, C.-W.~Chiang, G.~Cottin and D.~Shih,
Phys. Rev. D \textbf{101}, no. 5, 053001 (2020)
[arXiv:1908.08256 [hep-ph]].

\bibitem{Dreyer:2020brq}
F.~Dreyer and H.~Qu,
JHEP \textbf{03}, 052 (2021)
[arXiv:2012.08526 [hep-ph]].

\bibitem{Guo:2020vvt}
J.~Guo, J.~Li, T.~Li and R.~Zhang,
Phys. Rev. D \textbf{103}, no. 11, 116025 (2021)
[arXiv:2010.05464 [hep-ph]].

\bibitem{Gong:2022lye}
S.~Gong \emph{et al.},
JHEP \textbf{07}, 030 (2022)
[arXiv:2201.08187 [hep-ph]].

\bibitem{Larkoski:2017jix}
A.~J.~Larkoski, I.~Moult and B.~Nachman,
``Jet Substructure at the Large Hadron Collider: A Review of Recent Advances in Theory and Machine Learning,''
Phys. Rept. \textbf{841}, 1-63 (2020)
[arXiv:1709.04464 [hep-ph]].

\bibitem{Aguilar-Saavedra:2017rzt}
J.~A.~Aguilar-Saavedra and J.~H.~Collins and R.~K.~Mishra,
JHEP \textbf{11}, 163 (2017)
[arXiv:1709.01087 [hep-ph]].

\bibitem{Aguilar-Saavedra:2020sxp}
J.~A.~Aguilar-Saavedra and B.~Zald\'\i{}var,
Eur. Phys. J. C \textbf{80}, no. 6, 530 (2020)
[arXiv:2002.12320 [hep-ph]].

\bibitem{Aguilar-Saavedra:2020uhm}
J.~A.~Aguilar-Saavedra, F.~R.~Joaquim and J.~F.~Seabra,
JHEP \textbf{03}, 012 (2021)
[arXiv:2008.12792 [hep-ph]].

\bibitem{Aguilar-Saavedra:2021rjk}
J.~A.~Aguilar-Saavedra,
Eur. Phys. J. C \textbf{81}, no. 8, 734 (2021)
[arXiv:2102.01667 [hep-ph]].

\bibitem{Aguilar-Saavedra:2021utu}
J.~A.~Aguilar-Saavedra,
Eur. Phys. J. C \textbf{82}, no. 3, 270 (2022)
[arXiv:2111.02647 [hep-ph]].

\bibitem{Aguilar-Saavedra:2022ejy}
J.~A.~Aguilar-Saavedra,
Eur. Phys. J. C \textbf{82}, no. 2, 130 (2022)
[arXiv:2201.11143 [hep-ph]].

\bibitem{Cheng:2022gma}
T.~Cheng and A.~Courville,
JHEP \textbf{10}, 152 (2022)
[arXiv:2201.07199 [hep-ph]].


\bibitem{Metodiev:2017vrx}
E.~M.~Metodiev, B.~Nachman and J.~Thaler,
JHEP \textbf{10}, 174 (2017)
[arXiv:1708.02949 [hep-ph]].

\bibitem{Collins:2018epr}
J.~H.~Collins, K.~Howe and B.~Nachman,
Phys. Rev. Lett. \textbf{121}, no. 24, 241803 (2018)
[arXiv:1805.02664 [hep-ph]].

\bibitem{Collins:2019jip}
J.~H.~Collins, K.~Howe and B.~Nachman,
Phys. Rev. D \textbf{99}, no. 1, 014038 (2019)
[arXiv:1902.02634 [hep-ph]].

\bibitem{Heimel:2018mkt}
T.~Heimel, G.~Kasieczka, T.~Plehn and J.~M.~Thompson,
SciPost Phys. \textbf{6}, no. 3, 030 (2019)
[arXiv:1808.08979 [hep-ph]].

\bibitem{Farina:2018fyg}
M.~Farina, Y.~Nakai and D.~Shih,
Phys. Rev. D \textbf{101}, no. 7, 075021 (2020)
[arXiv:1808.08992 [hep-ph]].

\bibitem{Blance:2019ibf}
A.~Blance, M.~Spannowsky and P.~Waite,
JHEP \textbf{10}, 047 (2019)
[arXiv:1905.10384 [hep-ph]].

\bibitem{Hajer:2018kqm}
J.~Hajer, Y-Y.~Li, T.~Liu and H.~Wang,
Phys. Rev. D \textbf{101}, no. 7, 076015 (2020)
[arXiv:1807.10261 [hep-ph]].

\bibitem{Amram:2020ykb}
O.~Amram and C.~M.~Suarez,
JHEP \textbf{01}, 153 (2021)
[arXiv:2002.12376 [hep-ph]].

\bibitem{Bortolato:2021zic}
B.~Bortolato, A.~Smolkovi\v{c}, B.~M.~Dillon and J.~F.~Kamenik,
Phys. Rev. D \textbf{105}, no. 11, 115009 (2022)
[arXiv:2103.06595 [hep-ph]].

\bibitem{Cheng:2020dal}
T.~Cheng, J-F.~Arguin, J.~Leissner-Martin, J.~Pilette and T.~Golling
Phys. Rev. D \textbf{107}, no. 1, 016002 (2023)
[arXiv:2007.01850 [hep-ph]].

\bibitem{Nachman:2020lpy}
B.~Nachman and D.~Shih,
Phys. Rev. D \textbf{101}, 075042 (2020)
[arXiv:2001.04990 [hep-ph]].

\bibitem{Hallin:2021wme}
A.~Hallin \emph{et al.},
Phys. Rev. D \textbf{106}, no. 5, 055006 (2022)
[arXiv:2109.00546 [hep-ph]].

\bibitem{DeSimone:2018efk}
A.~De Simone and T.~Jacques,
Eur. Phys. J. C \textbf{79}, no. 4, 289 (2019)
[arXiv:1807.06038 [hep-ph]].

\bibitem{DAgnolo:2018cun}
R.~T.~D'Agnolo and A.~Wulzer,
Phys. Rev. D \textbf{99}, no. 1, 015014 (2019)
[arXiv:1806.02350 [hep-ph]].

\bibitem{Dillon:2019cqt}
B.~M.~Dillon, D.~A.~Faroughy and J.~F.~Kamenik,
Phys. Rev. D \textbf{100}, no. 5, 056002 (2019)
[arXiv:1904.04200 [hep-ph]].

\bibitem{Dillon:2020quc}
B.~M.~Dillon, D.~A.~Faroughy, J.~F.~Kamenik and M.~Szewc,
JHEP \textbf{10}, 206 (2020)
[arXiv:2005.12319 [hep-ph]].

\bibitem{Andreassen:2020nkr}
A.~Andreassen, B.~Nachman and D.~Shih,
Phys. Rev. D \textbf{101}, no. 9, 095004 (2020)
[arXiv:2001.05001 [hep-ph]].

\bibitem{Khosa:2020qrz}
C.~K.~Khosa and V.~Sanz,
[arXiv:2007.14462 [cs.LG]].







\bibitem{Chen:2016:XST:2939672.2939785}
T.~Chen and C.~Guestrin,
Proceedings of the 22nd ACM SIGKDD International Conference on Knowledge Discovery and Data Mining (KDD'16), 785 (2016)
[arXiv:1603.02754 [hep-ph]].

\bibitem{Adam-Bourdarios:2015pye}
C.~Adam-Bourdarios, G.~Cowan, C.~Germain-Renaud, I.~Guyon, B.~K\'egl and D.~Rousseau,
J. Phys. Conf. Ser. \textbf{82}, no. 7, 072015 (2015).

\bibitem{ATLAS:2017ztq}
M.~Aaboud \emph{et al.} [ATLAS],
Phys. Rev. D \textbf{97}, no. 7, 072003 (2018).
[arXiv:1712.08891 [hep-ex]].

\bibitem{CMS:2020tkr}
A.~Sirunyan \emph{et al.} [CMS],
JHEP \textbf{03}, 257 (2021).
[arXiv:2011.12373 [hep-ex]].

\bibitem{ATLAS:2021ifb}
G.~Aad \emph{et al.} [ATLAS],
Phys. Rev. D \textbf{106}, no. 5, 052001 (2022).
[arXiv:2112.11876 [hep-ex]].

\bibitem{CMS:2020cga}
A.~Sirunyan \emph{et al.} [CMS],
Phys. Rev. Lett. \textbf{125}, no. 6, 061801 (2020).
[arXiv:2003.10866 [hep-ex]].

\bibitem{CMS:2021sdq}
A.~Tumasyan \emph{et al.} [CMS],
JHEP \textbf{06}, 012 (2022).
[arXiv:2110.04836 [hep-ex]].

\bibitem{ATLAS:2021jbf}
G.~Aad \emph{et al.} [ATLAS],
JHEP \textbf{10}, 013 (2021).
[arXiv:2104.13240 [hep-ex]].

\bibitem{ATLAS:2020xov}
G.~Aad \emph{et al.} [ATLAS],
Eur. Phys. J. C \textbf{81}, no. 4, 313 (2021).
[arXiv:2010.02098 [hep-ex]].

\bibitem{CMS:2020kwy}
A.~Sirunyan \emph{et al.} [CMS],
JHEP \textbf{01}, 163 (2021).
[arXiv:2007.05658 [hep-ex]].

\bibitem{Odagiu:2024bkp}
P.~Odagiu \emph{et al.},
[arXiv:2402.01876 [hep-ex]].

\bibitem{Zabi:2020gjd}
A.~Zabi \textit{et al.} [CMS],
CERN-LHCC-2020-004.

\bibitem{Summers:2020xiy}
S.~Summers \emph{et al.},
JINST \textbf{15} 05, P05026 (2020).
[arXiv:2002.02534 [physics.comp-ph]].

\bibitem{CMSslides}
C.~Savard,
``Track quality machine learning models on FPGAs for the CMS Phase 2 Level 1 trigger'',
\emph{presented at} Fast Machine Learning for Science Workshop, 2020.

\bibitem{Alwall:2014hca}
J.~Alwall, R.~Frederix, S.~Frixione, V.~Hirschi, F.~Maltoni, O.~Mattelaer, H.~S.~Shao, T.~Stelzer, P.~Torrielli and M.~Zaro,
JHEP \textbf{07}, 079 (2014)
[arXiv:1405.0301 [hep-ph]].

\bibitem{protos}
J.A. Aguilar-Saavedra, Protos, a PROgram for TOp Simulations,
http://jaguilar.web. cern.ch/jaguilar/protos/

\bibitem{Sjostrand:2007gs}
T.~Sjostrand, S.~Mrenna and P.~Z.~Skands,
Comput. Phys. Commun. \textbf{178}, 852-867 (2008)
[arXiv:0710.3820 [hep-ph]].

\bibitem{deFavereau:2013fsa}
J.~de Favereau \textit{et al.} [DELPHES 3],
JHEP \textbf{02}, 057 (2014)
[arXiv:1307.6346 [hep-ex]].

\bibitem{Cacciari:2011ma}
M.~Cacciari, G.~P.~Salam and G.~Soyez,
Eur. Phys. J. C \textbf{72}, 1896 (2012)
[arXiv:1111.6097 [hep-ph]].

\bibitem{Cacciari:2008gp}
M.~Cacciari, G.~P.~Salam and G.~Soyez,
JHEP \textbf{04}, 063 (2008)
[arXiv:0802.1189 [hep-ph]].

\bibitem{Dreyer:2018tjj}
F.~A.~Dreyer, L.~Necib, G.~Soyez and J.~Thaler,
JHEP \textbf{06}, 093 (2018)
[arXiv:1804.03657 [hep-ph]].

\bibitem{Efron:526679}
B.~Efron and R. J. Tibshirani, \emph{An introduction to the bootstrap},
Chapman \& Hall/CRC monographs on statistics and applied (1993).



\end{thebibliography}
\end{document}